\documentclass[aps,pra,twocolumn,letterpaper,superscriptaddress,10pt]{revtex4-2}
\usepackage{amssymb,amsthm,amsmath,amsfonts}
\usepackage{graphicx,ulem}
\usepackage[pdftex,dvipsnames,usenames]{xcolor}
\usepackage[colorlinks=true,urlcolor=blue,citecolor=blue,linkcolor=blue]{hyperref}

\newcommand{\revision}[1]{\textcolor{black}{#1}}

\begin{document}

\title{\revision{Experimental retrieval of photon statistics from click detection}}

\author{S. Krishnaswamy}
    \affiliation{Paderborn University, Institute for Photonic Quantum Systems (PhoQS), Theoretical Quantum Science, Warburger Stra\ss{}e 100, 33098 Paderborn, Germany}

\author{F. Schlue}
    \affiliation{Paderborn University, Institute for Photonic Quantum Systems (PhoQS), Integrated Quantum Optics, Warburger Stra\ss{}e 100, 33098 Paderborn, Germany}

\author{L. Ares}
     \email{laurares@mail.uni-paderborn.de}
    \affiliation{Paderborn University, Institute for Photonic Quantum Systems (PhoQS), Theoretical Quantum Science, Warburger Stra\ss{}e 100, 33098 Paderborn, Germany}

\author{V. Dyachuk}
    \affiliation{Paderborn University, Institute for Photonic Quantum Systems (PhoQS), Integrated Quantum Optics, Warburger Stra\ss{}e 100, 33098 Paderborn, Germany}

\author{M. Stefszky}
    \affiliation{Paderborn University, Institute for Photonic Quantum Systems (PhoQS), Integrated Quantum Optics, Warburger Stra\ss{}e 100, 33098 Paderborn, Germany}

\author{B. Brecht}
    \affiliation{Paderborn University, Institute for Photonic Quantum Systems (PhoQS), Integrated Quantum Optics, Warburger Stra\ss{}e 100, 33098 Paderborn, Germany}

\author{C. Silberhorn}
    \affiliation{Paderborn University, Institute for Photonic Quantum Systems (PhoQS), Integrated Quantum Optics, Warburger Stra\ss{}e 100, 33098 Paderborn, Germany}

\author{J. Sperling}
    \affiliation{Paderborn University, Institute for Photonic Quantum Systems (PhoQS), Theoretical Quantum Science, Warburger Stra\ss{}e 100, 33098 Paderborn, Germany}

\date{\today}

\begin{abstract}
    We utilize click-counting theory for the reconstruction of photon statistics\revision{, employing an analytic pseudo-inversion method}.
    A reconfigurable time-bin multiplexing, click-counting detector is \revision{experimentally implemented, allowing for altering the photon-number resolution}.
    A detector tomography is carried out, yielding vital measurement features, such as quantum efficiencies and cross-talk rates.
    We gauge the success of the pseudo-inversion by applying the Mandel and binomial parameters, resulting in an additional interpretation of these parameters for the discrimination of distinct quantum statistics.
    For coherent states\revision{, which lie} at the classical-nonclassical boundary, both parameters are highly sensitive measures, probing the kind of statistics and the reconstruction performance.
    In addition, we apply a loss deconvolution technique to account for detection losses.
\end{abstract}

\maketitle

\section{Introduction}
\label{sec:Introduction}

    Since the beginning of quantum mechanics, the field of quantum optics has attracted an extraordinary interest for purposes of explaining optical phenomena that are outside the realm of classical physics \cite{E05,BT56,M59}.
    Since then, ongoing research has employed nonclassical optical states \cite{G63,M59} to develop and advance photonic quantum technologies \cite{KLM01,VBR08,BFV09}.
    For understanding and manipulating quantum light, a prerequisite is the ability to measure quantum-optical systems \cite{S07}.
    The information obtained from these measurements allows us to quantify the resourcefulness of quantum states by reconstructing their full interference structure \cite{SBRF93,ALP95,WV96}.
    One of the most crucial questions is determining whether a given light field is quantum \cite{M86,MVC19,GB20}, often investigated via photon statistics \cite{ATCR04,TPHMM24} for which photo-detectors are the main tool to provide this information.

    However, true photon-number-resolving detectors, which can measure arbitrarily high photon counts, are relatively inaccessible, despite ample research efforts.
    By contrast, detectors that record clicks as a result of any number of photons are readily available for on-demand usage due to their technological abundance \cite{GOCLSSVDWS01,CS07,SPMT13,ASY15,ZCLGESDZ21}.
    Therefore, it is of utmost importance to tailor strategies that account for the nature of click detection when characterizing nonclassical photon statistics.

    The quantum-optical click-detection theory allows one to formulate a proper model of click-counting systems, which may include up to 100 individual click detectors \cite{BDFL08,KASVSH17,TEBSS21,EHBAGDCP23,CZWSTT23}.
    To this end, experimental methods, such as multiplexing \cite{PTKJ96,ASSBW03,RHHPH03,SECMRKNLGWAV17}, are often used.
    An operator version of a binomial probability distribution arises as the general description of click-counting devices \cite{SVA12}.
    Thus, the earlier notion of sub-Poissonian light for photon-number statistics \cite{M82} is updated to the concept of sub-binomial light as its click-based equivalent \cite{SVA12a}, being further supplemented by the concepts of sub-Poisson-binomial light and sub-multinomial light \cite{LFPR16,SCEMRKNLGVAW17}.
    For identifying sub-binomial light, the binomial parameter $Q_B$ can be experimentally applied \cite{BDJDBW13,HSPGHNVS16}, just like the Mandel parameter $Q_M$ is for photon-number statistics.

    A manifold of sophisticated techniques to obtain photon-number distributions from measured data has been developed \cite{H97,ZABGGBRP05,YT05,ANAS09,DLCEPW09,SDGBRH12,CABHLRLS13,PLALKPDBG15,HDSJ19}.
    These reconstruction tools include maximum likelihood estimation \cite{B98}, detector tomography \cite{LFCPSREPW09}, and regularizations of ill-posed problems \cite{SSG09}, to name but a few.
    Another noteworthy method within the context of this paper is a detector calibration to determine the detector's response to light (see, e.g., Refs. \cite{CKS14,BKSSV17}), yielding important measurement characteristics, e.g., light-matter interactions \cite{WE92}, noise \cite{STG08}, and dead time and after pulses \cite{SSBSRVH24}.
    Regardless, the aforementioned methods commonly require a detection scheme with high efficiency, low noise, signals far from saturation, etc., being hardly satisfiable in practice.

    In this paper, we extract the photon statistics from measured click-counting statistics by utilizing an easy-to-implement pseudo-inversion technique \cite{KSVS18}, circumventing the complexities of other techniques.
    Furthermore, the Mandel and binomial parameters are employed unconventionally as discriminators between click-type and photon-number-type quantum statistics, assessing the success of the pseudo-inversion procedure.
    By applying coherent light\revision{, which lies} at the challenging classical-quantum boundary, our procedure removes a possible loophole which could lead to fake nonclassicality.
    A time-bin multiplexing detection system with superconducting single-photon nanowire detectors is realized in a modular manner, and a finely-grained detector calibration is carried out.
    Eventually, we also apply a loss deconvolution method to retrieve a near-ideal photon-number distribution.

    The remainder of the paper is organized as follows:
    The distinct photon-number and click-counting theories and their discrimination via $Q$ parameters are laid out in Sec. \ref{sec:TheoryFundamentals}.
    In Sec. \ref{sec:DetectorTomography}, we describe our experiment and present the reconstruction of the detector response function via coherent-state measurements.
    Section \ref{sec:PseudoInversion} includes our key results from the pseudo-inversion and the application of $Q$ parameters.
    The losses are treated in Sec. \ref{sec:LossDeconvolution}.
    Eventually, we conclude in Sec. \ref{sec:Conclusion}.

\section{Detection theories and quantum-statistical signatures}
\label{sec:TheoryFundamentals}
%


    In photoelectric detection theory \cite{KK64,MSW64}, ideal photodetectors can be represented through their positive operator-valued measure,
    $
    	\hat\Pi_n=|n\rangle\langle n|={:}e^{-\hat n}\hat n^n/n!{:}
    $,
    with $n\in\mathbb N$, the photon-number operator $\hat n$, and ${:}\cdots{:}$ denoting the normal-ordering prescription.
    For example, the photon-number distribution of coherent light is given by $p_n=\langle\alpha|\hat\Pi_n|\alpha\rangle=e^{-|\alpha|^2}|\alpha|^{2n}/n!$, with a coherent amplitude $\alpha$.
    To account for imperfections, $\hat n$ can be replaced by a general response function $\hat\Gamma$ \cite{KK64}, e.g.,
    $
    	\hat\Gamma=\eta\hat n+\nu\hat1
    $,
    where $\eta$ is the quantum efficiency and $\nu$ describes dark counts.
    Photon counting for general response functions $\hat\Gamma$ always takes the form of quantum-Poissonian distribution.
    To describe the quantum-statistical properties of states of light, access to the photon-number distribution is essential, which is difficult because of a lack of true photon-number-resolving detectors.

\begin{figure}
    \includegraphics[width=0.3\textwidth]{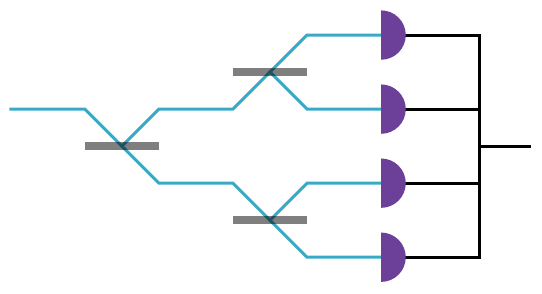}
    \\
    \includegraphics[width=0.3\textwidth]{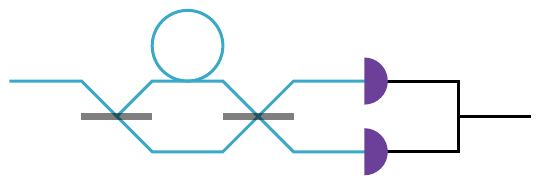}
    \includegraphics[width=0.325\textwidth]{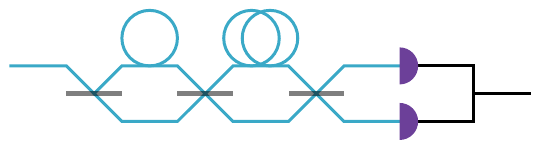}
    \caption{%
        The click-counting theory applies to, for example, the three depicted multiplexing schemes, in which light is uniformly split spatially ($50{:}50$ beam splitters) and temporally (delay loops), before being measured with on-off click detectors that collectively yield a total number of clicks.
        The top, spatial multiplexing layout utilizes three beam splitters and $N=4$ detectors.
        The middle configuration uses one less beam splitter and only two detectors.
        Together with the delay line, this results in early and late time bins in the top and bottom detector, i.e., $N=4$ detection bins.
        The bottom scheme doubles the number of time bins by introducing only one extra beam splitter and a delay double the length of the prior loop, resulting in a $N=8$ time-bin multiplexing detector.
        The latter two schemes, being more efficient than the spatial multiplexing in the number of optical elements, are experimentally realized in this paper.
    }\label{fig:MultiplexingSchemes}
\end{figure}

    Nowadays, common detector technologies employ so-called click detectors where a ``click'' is recorded if one or an arbitrary number of photons are absorbed, and ``no click''' is recorded otherwise \cite{GOCLSSVDWS01,CS07,ASY15,ZCLGESDZ21}.
    Furthermore, incident light may be uniformly distributed across $N$ click detectors, commonly referred to as multiplexing \cite{PTKJ96,RHHPH03,ASSBW03};
    see Fig. \ref{fig:MultiplexingSchemes}.
    Then, we can express realistic click detectors through \cite{SVA12}
    \begin{equation}
        \label{eq:POVM}
		\hat\pi_k={:}\binom{N}{k}\left(e^{-\hat \Gamma}\right)^{N-k}\left(\hat 1-e^{-\hat \Gamma}\right)^k{:},
	\end{equation}
	for $k\in\{0,\ldots,N\}$ total clicks.
	One strategy of multiplexing is time-bin multiplexing, which is the detection scenario implemented in our experiment (two bottom schemes in Fig. \ref{fig:MultiplexingSchemes}).
    In contrast to the photon-number distribution, the click-counting distribution [Eq. \eqref{eq:POVM}] is a quantum-operator analog to a binomial distribution.
    For example, the lossless and dark-count-free click-counting distribution for coherent light is $c_k=\langle\alpha|\hat\pi_k|\alpha\rangle=\binom{N}{k}\big(e^{-|\alpha|^2/N}\big)^{N-k}\big(1-e^{-|\alpha|^2/N}\big)^{k}$.

    It is important to emphasize that $k$ clicks and $n$ photons are related, but dissimilar concepts (even if $k=n$, $\eta=1$, and $\nu=0$), which is discussed later in more detail.


    One can use the Poissonian behavior in the photoelectric detection scenario as a reference for probing nonclassicality of light \cite{M82}.
    Analogously, one can characterize the super- and sub-binomial behavior with click detection systems for determining the quantumness of the signal in theory and experiment \cite{SVA12a,BDJDBW13}.
    In photoelectric measurements, the Mandel parameter is utilized,
    \begin{equation}
        \label{eq:QMandel}
        Q_M= \frac{
            \overline{(\Delta n)^2}
        }{
            \overline{n}
        } - 1,
    \end{equation}
    with the expected photon number $\overline n=\sum_{n=0}^\infty n p_n$ and variance $\overline{(\Delta n)^2}=\overline{n^2}-\overline{n}^2$ of the distribution $(p_n)_{n\in\mathbb N}$.
    A value $Q_M=0$ describes a Poisson distribution,
    classical light obeys $Q_M\geq 0$,
    and nonclassicality is verified when $Q_M<0$.
    In the same manner, we utilize the binomial parameter $Q_B$ for click-counting devices \cite{SVA12a},
    \begin{equation}
        \label{eq:QBinomial}
        Q_B= N\frac{\overline{(\Delta k)^2}}{\overline{k}\left(N-\overline k\right)}-1,
    \end{equation}
    using the variance $\overline{(\Delta k)^2}$ and expected click number $\overline k$ of the measured click-counting distribution $(c_k)_{k\in\{0,\ldots,N\}}$;
    $Q_B=0$ is achieved for a perfect binomial distribution.

\begin{figure}
    \includegraphics[width=.38\textwidth]{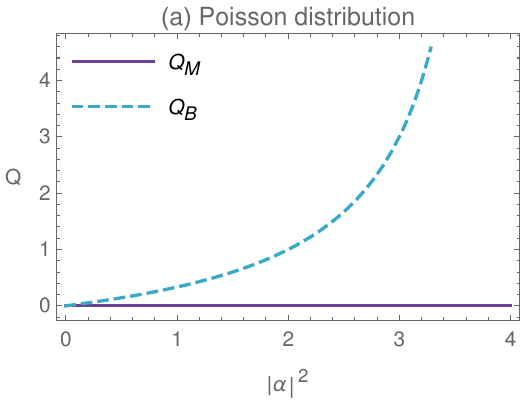}
    \\[2ex]
	\includegraphics[width=.4\textwidth]{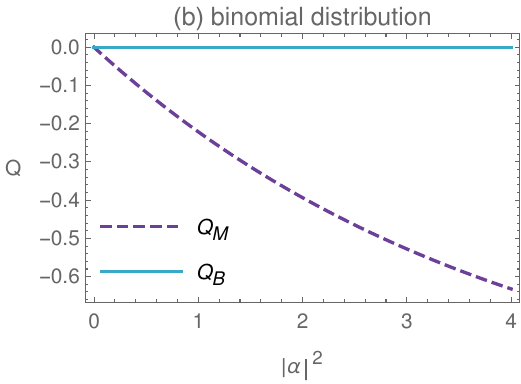}
	\caption{%
        $Q$ parameters applied to binomial ($N=4$) and Poisson probability distributions.
        The top plot (a) depicts the result for the photon-number distribution of a coherent state as a function of a mean photon number $|\alpha|^2$.
        $Q_M=0$ (solid, purple) shows the Poisson nature of the photon-number statistics,
        and $Q_B\neq0$ (dashed, blue) certifies incompatibility with binomial statistics.
        The bottom plot (b) analogously gives both $Q$ parameters when applied to a click-counting distribution of a coherent state.
        $Q_M\neq 0$ (dashed, purple) and $Q_B=0$ (solid, blue) yield the expected falsification and compatibility with Poissonian and binomial behaviors, respectively.
    }\label{fig:theoryQs}
\end{figure}

    Introducing the parameter $Q_B$ is necessary because if one applies $Q_M$ to click-based measurement devices, we may get a false indication of nonclassicality \cite{SVA12,SVA12a}.

    Here, we additionally attribute another function to $Q_B$ and $Q_M$ as they enable us to determine the Poissonian or binomial behavior of a detection system and the character of reconstructed distributions from data, Fig. \ref{fig:theoryQs}.
    In general, $Q_{B/M}<0$ and $Q_{B/M}>0$ means that the distribution under consideration is, respectively, too narrow and too broad for a binomial or Poissonian distribution.
    Also, coherent states, as employed in our experiment, are at the classical-quantum boundary, exactly yielding $Q_B=0$ and $Q_M=0$.
    Thus, such edge states present a highly sensitive testbed for our method, for example, to ensure that no fake quantum effects are introduced via the reconstruction procedure.

\section{Setup and detector calibration}
\label{sec:DetectorTomography}

    In this section, we describe and characterize our experiment.
    The detector characteristics of the whole measurement system are analyzed through a thorough detector tomography.
    We also probe the cross-talk and splitting uniformity between the detection time-bins.

\begin{figure}
    \includegraphics[width=.45\textwidth]{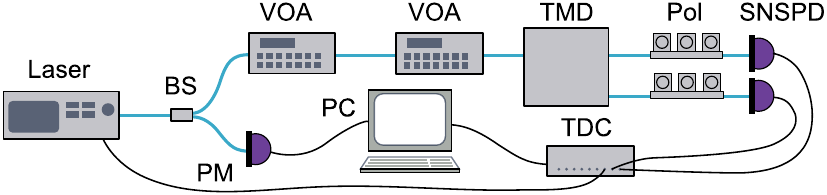}
	\caption{%
        A variable fiber beam splitter (BS) is used to divide a beam into two paths, one of which is detected with a power meter (PM), and the other with a superconducting nanowire single-photon detector (SNSPD) after passing through two variable optical attenuators (VOA) and the time-multiplexed detector (TMD) system.
        The clicks are recorded with a time-to-digital converter (TDC).
        Full description can be found in the text.
    }\label{fig:ExperimentSetup}
\end{figure}

\subsection{Setup description}

    We begin with a description of the realized detection system;
    see Fig. \ref{fig:ExperimentSetup}.
    We use a pulsed laser with a pulse length of $100\,\mathrm{ps}$ at $\lambda=1549.8\,\mathrm{nm}$ wavelength as a coherent light source.
    The repetition rate of the pulsed laser is $r=2\,\mathrm{MHz}$.
    The pulse intensity is split via a beam splitter.
    The bright output, containing approximately $94\%$ of the light, is directly measured with a power meter and is used as a reference to monitor power drifts.
    The remaining $6\%$ is attenuated to the single-photon level with two variable optical attenuators.
    The two attenuators have a rated error of $\pm0.3\,\mathrm{dB}$ and are individually characterized for linearity.
    The attenuated coherent laser pulses are connected to the time-bin multiplexing scheme, which can be set to consist of either $N=4$ or $8$ detection bins.
    The output from the time multiplexing is detected with two superconducting nanowire single-photon detectors.
    A time-to-digital converter is used to log the arrival time of the detected clicks relative to the laser pulses.

    Before the measurements are performed, the input photon flux to the detectors is calibrated.
    For that, the power exiting the two output ports of the variable beam splitter is measured with the power meter.
    Additionally, the zero attenuation of the attenuators is measured and is added as a constant offset to the selected attenuation.
    Together with the constant power monitoring of the input power, this calibration provides the photon flux entering the time-bin multiplexing detection system within the measurement uncertainty.

    With the input photon-number statistics defined, the next task is to define the relevant time bins such that the click statistics can be correctly measured.
    To this end, we calibrate the output time bins in each detector to the time reference of the laser.
    This allows narrow time filtering, here $1\,\mathrm{ns}$, around the expected time bins.
    The system then records the measured click number for each trigger provided by the laser, as well as the click pattern of the time bins.

\subsection{Characterization}

\paragraph{Photon-number estimation.}

    We measure the laser power $P_0=1.5\times10^{-5}\,\mathrm{W}$ with the power meter and carefully tally all losses in our setup before the attenuated laser light enters the time-bin multiplexing detector, $L_0=33.4\,\mathrm{dB}$.
    In addition, the variable attenuators ($55\,\mathrm{dB}\leq L\leq 70\,\mathrm{dB}$) yield coherent states with different mean photon numbers $\bar n=|\alpha|^2$.
    Together, this leads to a power at the input of the detector of $P=10^{-(L_0+L)/10}P_0$.
    Using the relations
    \begin{equation}
    	P=\bar nrhf
    	\quad\leftrightarrow\quad
    	\bar n=\frac{10^{-(L_0+L)/10}P_0\lambda}{rhc},
    \end{equation}
    with the central wavelength $c/f=\lambda$, the repetition rate $r$, the Plank constant $h$, and the speed of light $c$, we obtain mean photon numbers
    \begin{equation}
    	0.027\leq\bar n\leq 0.85
    \end{equation}
    at the input of the click-counting device.
    In this range, we measure sixteen and six attenuated coherent states for the four-bin and eight-bin configuration, respectively.
    Also note that the mean photon numbers are determined independently from the detection system to be characterized in the following.

\begin{figure}
    \includegraphics[width=.35\textwidth]{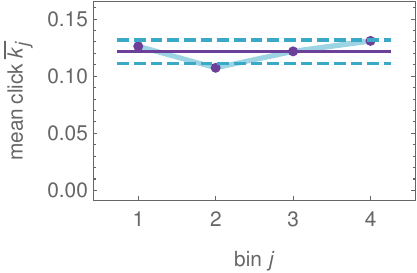}
    \\[2ex]
    \includegraphics[width=.4\textwidth]{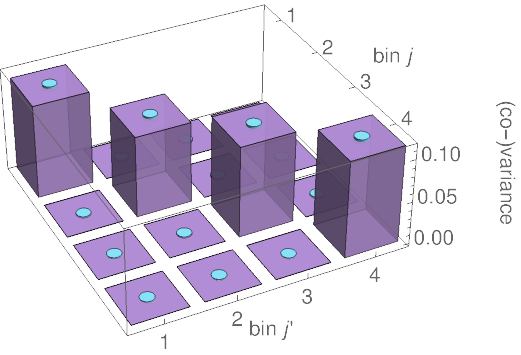}
	\caption{%
        The top plot depicts the mean clicks per bin, Eq. \eqref{eq:MeanEachBin}, depicted here for the $N=4$ configuration and the maximal photon number $\bar n=|\alpha|^2=0.84$.
        This distribution is compatible with a uniform distribution (purple, solid line);
        one standard deviation $\sigma$ of $\pm0.011$ clicks (blue, dashed lines) is shown.
        The bottom plot depicts bin-bin cross talk through vanishing off-diagonal covariances, Eq. \eqref{eq:BinBinCovar}, for the same data set as used above.
        No two bins $j\neq j'$ exhibit nonzero correlations within the margin of error (blue cylinder depicts $10\sigma$ for improved visibility).
        Similar results are found for all measured cases, including the $N=8$-bin scenario.
	}\label{fig:BinByBin}
\end{figure}

\paragraph{Bin uniformity and cross talk.}

    We can now analyze the detection system.
    We begin with observing the system's quality across individual detection bins.
    To this end, clicks $c_{k_1,\ldots,k_N}$ were recorded with $k_j=0$ and $1$ denoting no click or a click from the $j$th time bin.
    One measure for how well our system operates is the uniform distribution of light across detection bins, which can be quantitatively assessed with
    \begin{equation}
        \label{eq:MeanEachBin}
    	\overline{k_j}=\sum_{k_1,\ldots,k_N=0}^1 c_{k_1,\ldots,k_N} k_j,
    \end{equation}
    which is expected to be constant with respect to $j$ in case of a uniform distribution.
    The result for one data set is depicted in Fig. \ref{fig:BinByBin} (top).
    While deviations from perfect uniformity exist, they do not amount to significant biases across bins.

    Exceeding the performance of common characterizations, our bin-by-bin analysis also allows us to determine cross-talk between detection bins, here quantifying the temporal separation of time bins.
    This can be achieved by covariances
    \begin{equation}
        \label{eq:BinBinCovar}
    	\overline{(\Delta k_{j})(\Delta k_{j'})}
    	=\overline{k_{j}k_{j'}}-\overline{k_{j}}\,\overline{k_{j'}},
    \end{equation}
    which ought to be zero for $j\neq j'$, i.e., uncorrelated for coherent light.
    Exemplified again for one data set, Fig. \ref{fig:BinByBin} (bottom) shows that cross-correlations are insignificant, proving an excellent temporal separation.

\paragraph{Detector response.}

    We now transition from the bin-by-bin characterization to the characterization of the full detection system by virtue of the click-counting statistics $c_k=\sum_{k_1+\cdots+k_N=k}c_{k_1,\ldots,k_N}$, where $k\in\{0,\ldots,N\}$ denotes the total number of clicks.
    Inferring photon-number features from click detection gives us a strong motivation to carry out a detector calibration for obtaining the click detector's response function \cite{BKSSV17}.
    For this purpose, we recall that the first moment of the click-counting distribution reads \cite{SVA13,SBVHBAS15}
    \begin{equation}
        \label{eq:firstMoment}
        \overline{k}
        =N\left(
            1-\langle{:} e^{-\hat \Gamma}{:}\rangle
        \right)
        \quad\leftrightarrow\quad
        \Gamma=-\ln\left(1-\frac{\overline k}{N}\right).
    \end{equation}
    Therein, we used the simplification $\langle \alpha|{:}e^{-\hat \Gamma}{:}|\alpha\rangle=e^{-\Gamma}$ for normally ordered expectation values with coherent states.
    Furthermore, the assumed response function is
    \begin{equation}
        \label{eq:NonlinResponse}
            \hat \Gamma=\nu\hat 1+\eta\frac{\hat n}{N}
            +\gamma\left(\frac{\hat n}{N}\right)^2,
    \end{equation}
    where the scaling $\hat n/N$ accounts for the fact that input photons are distributed across the $N$ detection bins and the second-order term proportional to $\gamma$ probes the presence of nonlinear absorption properties \cite{SVA13,SECMRKNLGWAV17}.
    Note that we probed different orders of the Taylor approximation in the variable $\hat n/N$ and found that a linear response function suffices for our data.
    However, as a concrete example, we here focus on the second-order nonlinear case.
    Together with Eq. \eqref{eq:firstMoment}, we have the function
    \begin{equation}
        \label{eq:FitFunction}
    	|\alpha|^2=\bar n\,\mapsto\,\Gamma=\nu+\eta\frac{\bar n}{N}+\gamma\left(\frac{\bar n}{N}\right)^2,
    \end{equation}
    that can be fitted to our data.

\begin{figure}
	\includegraphics[width=.35\textwidth]{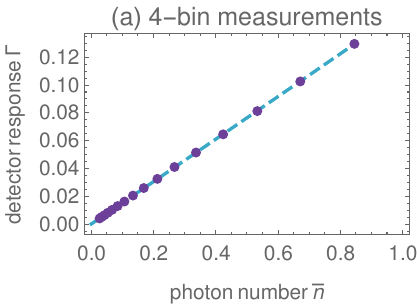}
	\\[4ex]
	\includegraphics[width=.35\textwidth]{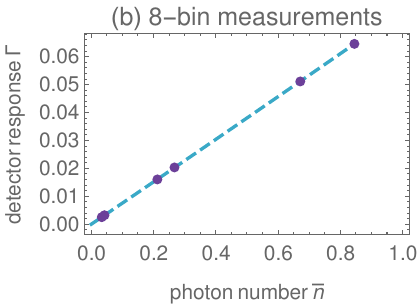}
	\caption{%
        Determination of the response function [Eq. \eqref{eq:FitFunction}; blue, dashed lines] from measured mean click numbers [Eq. \eqref{eq:firstMoment}; thick, purple points] for $N=4$ and $8$ time-bin detection setups in the top and bottom depiction, respectively.
        In both cases, the response is governed by a linear behavior, with quantum efficiencies $\eta|_{N=4}=60.8\%$ and $\eta|_{N=8}=60.5\%$.
        Dark counts (less than $2\times 10^{-4}$) and nonlinear contributions ($\gamma\bar n^2/N^2<10^{-3}$) are negligible in both cases.
        For $N=4$ and $8$, the coefficient of determination is $R^2>0.9999$, quantitatively showing a near-perfect agreement between data and fit.
	}\label{fig:Calibration}
\end{figure}

    We use a nonlinear regression in Fig. \ref{fig:Calibration} to retrieve the coefficients in Eq. \eqref{eq:FitFunction}.
    For both detector variants, $N=4$ and $8$, the dark count contributions are negligible, $\nu|_{N=4}=1.8\times 10^{-4}(1\pm0.4)$ and $\nu|_{N=8}=5.8\times10^{-6}(1\pm8)$.
    Also, the nonlinear contribution that scales with $(\bar n/N)^2$ is small, $\gamma|_{N=4}=1.9\times 10^{-2}(1\pm0.6)$ and $\gamma|_{N=8}=6.2\times 10^{-2}(1\pm0.5)$.
    (None of these parameters is different from zero with statistical significance, a three-standard-deviation error margin.)
    Most importantly, the quantum efficiency of the click-counting device is about $60\%$, $\eta|_{N=4}=60.8\%(1\pm0.004)$ and $\eta|_{N=8}=60.5\%(1\pm0.005)$, including both optical losses in the multiplexing stage and non-unit efficiencies of individual on-off detectors.

\paragraph{Discussion.}

    Our comprehensive detector characterization certifies an excellent performance of the whole click detection system.
    The response function is mostly linear, exhibiting insignificant dark count contributions and detection nonlinearities.
    The overall quantum efficiency of the click-counting devices is estimated as $60\%$.
    Furthermore, we also ascertain the quality of the setup in terms of the uniformity of the distribution of photons across time bins without overlap, i.e., no cross talk.

\section{Pseudo-photon-number reconstruction}
\label{sec:PseudoInversion}

    In contrast to previous studies, we exploit the binomial click statistics in Eq. \eqref{eq:POVM} and implement a theoretically proposed pseudo-inversion approach \cite{KSVS18}.
    The click-counting distribution can be expressed through the photoelectric statistics for a quantum efficiency $\eta$ as \cite{KSVS18}
    \begin{equation}
        \label{eq:LossyPhotonsToClicks}
        c_k=\sum_{n=0}^\infty\binom{N}{k}\frac{k!}{N^n}\begin{Bmatrix}n\\k
        \end{Bmatrix} p_n(\eta) ,
    \end{equation}
    where $k\in\{0,\ldots, N\}$ is the number of clicks, $N$ is the total number of detection bins, and $\left\lbrace\begin{smallmatrix}n\\k \end{smallmatrix}\right\rbrace$ are the Stirling numbers of the second kind.
    In vector notation, $\vec c=(c_k)_{k\in\{0,\ldots,N\}}$ and $\vec p(\eta)=(p_n(\eta))_{n\in\mathbb N}$, Eq. \eqref{eq:LossyPhotonsToClicks} takes the following form:
    \begin{equation}
        \label{eq:LossyPhotonsToClicksVectorForm}
    \begin{aligned}
        \vec c=\boldsymbol{C}\vec p(\eta),
        \quad\text{where}\quad
        \boldsymbol C=(C_{k,n})_{k\in\{0,\ldots,N\},n\in\mathbb N}
        \\
        \text{and}\quad
        C_{k,n}=\binom{N}{k}\frac{k!}{N^n}\begin{Bmatrix}n\\k
        \end{Bmatrix}.
    \end{aligned}
    \end{equation}

\begin{figure}
	\includegraphics[width=.4\textwidth]{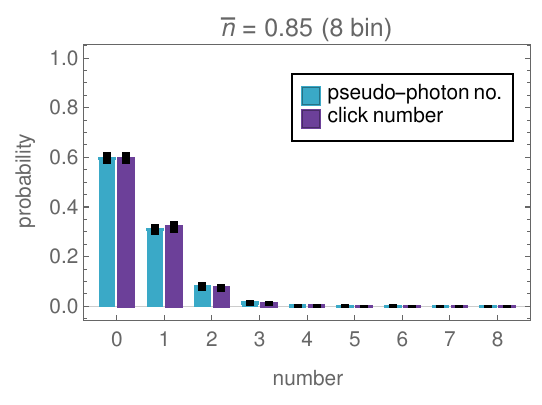}
	\caption{%
        Example of reconstructed [Eq. \eqref{eq:PseudoInversion}] pseudo-photon-number distribution $p'_n(\eta)$ (blue, left) and measured click-counting distribution $c_k$ (purple, right), with $k,n\in\{0,\ldots,N\}$ for $N=8$.
        (For visibility, a $50\sigma$ error bar is shown in black.)
        A direct comparison of the measured and reconstructed probability distributions would lead to the conclusion that they are rather similar.
        However, this superficial belief is disproved in Fig. \ref{fig:ExperimentQParameters} where we apply $Q$ parameters to distinguish dissimilar quantum statistics.
	}\label{fig:ClickAndPseudoPhotonNo}
\end{figure}

\begin{figure*}
    \includegraphics[width=.4\textwidth]{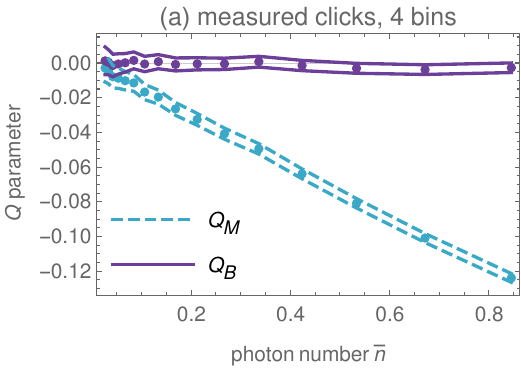}
    \qquad
    \includegraphics[width=.4\textwidth]{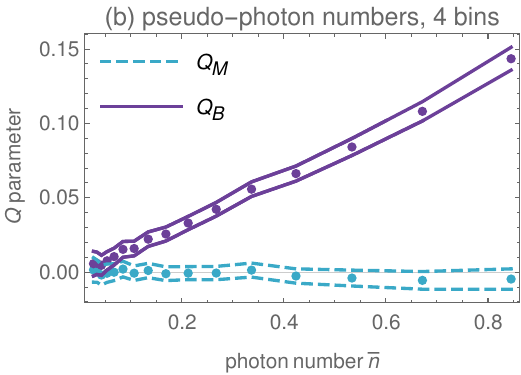}
    \\[2ex]
    \includegraphics[width=.4\textwidth]{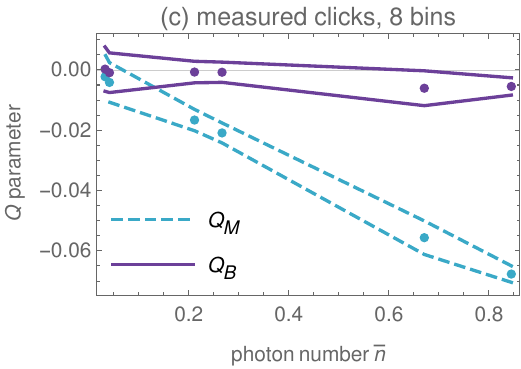}
    \qquad
    \includegraphics[width=.4\textwidth]{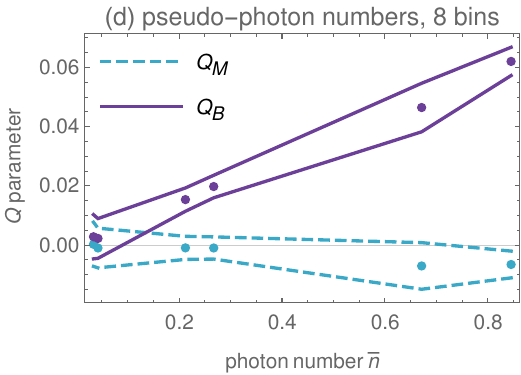}
	\caption{%
        Application of the Mandel $Q_M$ parameter [Eq. \eqref{eq:QMandel}] and binomial $Q_B$ parameter [Eq. \eqref{eq:QBinomial}] for verifying and falsifying Poisson and binomial statistics, for $N=4$ [plots (a) and (b)] and $N=8$ detection bins [plots (c) and (d)], including a $1\sigma$ uncertainty (blue, dashed for $Q_M$ and purple, solid for $Q_B$).
        Plots (a) and (c) demonstrate that the measured click-counting data are compatible with a binomial distribution but distinct from a Poissonian photon-number distribution, as expected for coherent light.
        After pseudo-inversion, plots (b) and (d), the reconstructed statistics follow Poisson-like distributions, which are clearly distinct from binomial distributions, demonstrating the successful conversion from click-counting data to photon-number information.
	}\label{fig:ExperimentQParameters}
\end{figure*}

    The matrix $\boldsymbol C$ maps the infinite-dimensional photon number $\vec p(\eta)$ to the $(N+1)$-dimensional vector $\vec c$ of clicks.
    Therefore, it is an impossibility to invert the matrix $\boldsymbol C$ as an insufficient amount of information is available.
    Instead, we apply the pseudo-inverse $\boldsymbol C^+$ \cite{KSVS18}, which is of the same size as the click statistics, with matrix entries
    \begin{equation}
        \label{eq:PseudoInverse}
        C_{k,m}^{+}=\binom{N}{k}^{-1}\frac{N^m}{k!} \genfrac{[}{]}{0pt}{}{k}{m},
    \end{equation}
    for $m,k\in\{0,\ldots, N\}$ and with $\left[\begin{smallmatrix}k\\m \end{smallmatrix}\right]$ denoting the (signed) Stirling numbers of the first kind.
    In the context of Eq. \eqref{eq:PseudoInverse}, we refer to $m$ as the pseudo-photon number because it is obtained via the pseudo-inverse $\boldsymbol C^{+}$ that yields
    \begin{equation}
        \label{eq:PseudoInversion}
        \boldsymbol C^{+}\vec c=\vec{p}^{\,\prime}(\eta).
    \end{equation}
    Note that, for states with photon-number distributions with zero probability for photon numbers $n>N$, the reconstructed statistics is exact \cite{KSVS18}, i.e., $p_n^{\prime}(\eta)=p_n(\eta)$.
    Furthermore, no correction for losses has been made here;
    a loss deconvolution is carried out in Sec. \ref{sec:LossDeconvolution}.

    We apply the pseudo-inversion to our measured click statistics and compare the initial and resulting distribution in Fig. \ref{fig:ClickAndPseudoPhotonNo}.
    On a superficial level, one might think that the inversion does nothing substantial.
    However, upon closer inspection by probing the binomial and Poissonian character in Fig. \ref{fig:ExperimentQParameters}, we clearly find a distinction between the measured click distribution and the pseudo-photon-number distribution, which becomes increasingly prominent with increasing signal intensity of the coherent input state.
    In particular, we can infer that the measured data follow the expected binomial behavior [$Q_B=0$ for $\vec c\,$], and our extracted pseudo-photon values behave like a true Poisson statistics [$Q_M=0$ for $\vec p^{\,\prime}(\eta)$].
    The binomial and Mandel parameter are commonly used to probe the nonclassical behavior of light.
    Here, we assign an additional value to such parameters: a highly sensitive and effective measure to distinguish quantum-statistical signatures.
    
\section{Loss deconvolution}
\label{sec:LossDeconvolution}

    Thus far, we implemented the pseudo-inversion from the measured click-counting statistics $\vec c$ to the lossy pseudo-photon-number distribution $\vec p^{\,\prime}(\eta)$.
    Now, we additionally address the problem of removing attenuation for estimating the lossless photon-number distribution $\vec p^{\,\prime}=\vec p^{\,\prime}(1)$  \cite{STG08,SSG09}.
    Recall that we previously determined the efficiency $\eta=60\%$ in Sec. \ref{sec:DetectorTomography}.
    
\begin{figure*}
	\includegraphics[width=.7\textwidth]{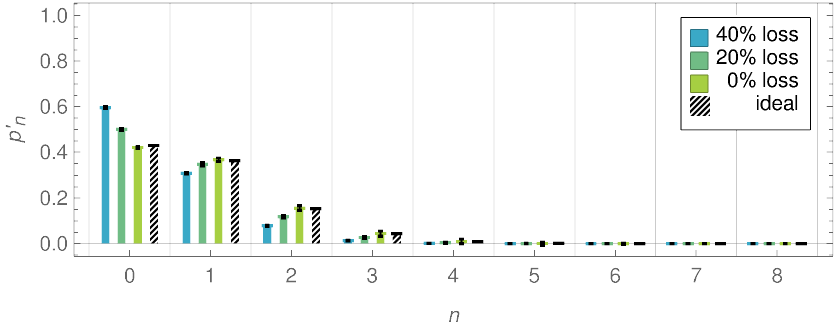}
	\caption{%
       Application of the loss deconvolution to the pseudo-photon-number distribution from Fig. \ref{fig:ClickAndPseudoPhotonNo}, with $\bar n=|\alpha|^ 2=0.85$ and $N=8$.
        The leftmost (blue), left-center (blue-green), and right-center (green) distributions correspond to the unaltered distribution $\vec p^{\,\prime}(60\%)$, the distribution $\vec p^{\,\prime}(80\%)$ in which half the experimental losses have been removed, and the lossless distribution $\vec p^{\,\prime}(100\%)$, respectively.
        For comparison, the rightmost (hatch shaded) distribution represents the expected, ideal distribution $p_n=e^{-|\alpha|^2}|\alpha|^{2n}/n!$ for coherent light and perfect detectors.
        In all cases of losses, the statistical consistency with a Poissonian distribution is preserved;
        i.e.,
        $Q_M|_{\eta=60\%}=-0.0065(1\pm70\%)$,
        $Q_M|_{\eta=80\%}=-0.0087(1\pm140\%)$,
        and $Q_M|_{\eta=100\%}=-0.011(1\pm360\%)$ are zero within an at most $1.5\sigma$ error margin (likewise, all negativities are insignificant).
	}\label{fig:LossDeconvolution}
\end{figure*}

    Beginning with the true photon statistics $\vec p$, the distribution undergoes the well-known loss-channel description, after which it gets convoluted into experimentally measured $\vec c$,
    \begin{equation}
    	\label{eq:BothConvolutions}
    \begin{aligned}
        &\quad
    	\vec p
    	\quad\mapsto\quad
    	\vec p(\eta)=\boldsymbol H(\eta)\vec p
    	\\
    	\quad\mapsto&\quad
    	\vec c=\boldsymbol C\vec p(\eta)=\boldsymbol C\boldsymbol H(\eta)\vec p,
    \end{aligned}
    \end{equation}
    where the linear loss map $\boldsymbol H(\eta)$ is defined through the matrix elements
    \begin{equation}
        \label{eq:LossMatrixEntries}
    	H_{n,m}(\eta)=\begin{cases}
            \binom{m}{n}\eta^n(1-\eta)^{m-n}
            & \text{if } m\geq n,\\
            0
            & \text{otherwise.}
        \end{cases}
    \end{equation}
    In terms of vector and matrix entries, this means $p_n(\eta)=\sum_{m=n}^\infty\binom{m}{n}\eta^n(1-\eta)^{m-n}p_n$, which is the common loss representation.
    The above matrix-valued function $\boldsymbol H$ satisfies the following properties:
    $\boldsymbol H(\eta_1)\boldsymbol H(\eta_2) =\boldsymbol H(\eta_1\eta_2)$ and $\boldsymbol H(1)$ is the identity.
    This further implies $\boldsymbol H(\eta)^{-1}=\boldsymbol H(1/\eta)$, which is useful for loss removal.

    Via the pseudo-inverse $\boldsymbol C^{+}$ in Eq. \eqref{eq:PseudoInversion}, we already carried out the crucial step in the $(N+1)$-photon subspace, resulting in $\vec p^{\,\prime}(\eta)$.
    According to the inversion of the order of maps in Eq. \eqref{eq:BothConvolutions}, the final step, on which we focus here, is a more common loss deconvolution.
    The matrix entries from Eq. \eqref{eq:LossMatrixEntries} for the needed inverse loss matrix read
    $
        H_{n,m}(1/\eta)=\binom{m}{n}(1/\eta)^n\left(1-1/\eta\right)^{m-n},
    $
    including arbitrarily large, negative contributions because of $1-1/\eta\to-\infty$ for $\eta\to 0$, rendering the deconvolution unstable \cite{SSG09}.
    This includes unphysical cases where loss-deconvoluted photon-numbers become negative, thus requiring some extra attention when removing losses from any data in any measurement scenario.

    Fig. \ref{fig:LossDeconvolution} shows our results for the deconvolution of losses from the pseudo-photon-number distribution obtained in Sec. \ref{sec:PseudoInversion}, where $\eta=60\%$.
    To show the progression to full loss removal, $\eta=100\%$, a half-way step with the efficiency $80\%$ is depicted together with the theoretically predicted, lossless Poisson distribution (black-white).
    From $Q_M\approx 0$, as discussed in the caption of Fig. \ref{fig:LossDeconvolution}, we can conclude that the Poissonian nature of our initial pseudo-inversion was successfully preserved by the loss deconvolution.
    In addition, our final, lossless quasi-photon-number distributions are in excellent agreement with the ideal photon-number measurement.

\section{Conclusion}
\label{sec:Conclusion}

    In this paper, a pseudo-inversion of click-counting data was carried out to obtain the otherwise inaccessible photon-number distribution.
    The Mandel and binomial parameters were proposed and used as measures to assess the quality of the reconstruction procedure, and to generally discern different types of quantum statistics.

    Our experimental implementation of the methodology is based on a custom-made, time-bin multiplexing detector that is easily configurable to a desired resolution.
    Based on detector tomography with coherent light, we analyzed the detection system's response, allowing us to determine essential properties, such as detection efficiency and cross-talk between time bins.
    To demonstrate the variable nature of our setup, both a four-bin and eight-bin configuration were analyzed in this manner.
    Using the binomial and Mandel parameter, we proved that our coherent-light data are consistent with a binomial click distribution and inconsistent with a Poissonian photon distribution as one would expect from the comparison of click and photoelectric counting theories.

    A pseudo-inversion approach was implemented to recover a pseudo-photon-number resolution up to four and eight photons, depending on the detector's configuration.
    While the resulting and initially measured distribution appear rather similar when plotted together, the Mandel and binomial parameters reveal the now Poissonian nature of the reconstructed distribution and rule out the previous binomial behavior, respectively.
    Thereby, the success of the pseudo-inversion as well as the effectiveness of the discrimination via the two parameters is demonstrated.
    Importantly, the pseudo-inversion is based on an analytically obtained matrix that is applied only once to each data set, circumventing difficulties and computational complexities of other competing reconstruction techniques.
    The reconstruction method was applied to coherent states at the boundary between classical and nonclassical light, presenting a uniquely challenging scenario.
    However, it is worth mentioning that the pseudo-inversion applies to general, unknown states.

    Eventually, a loss deconvolution was applied to correct for the previously estimated losses of the detection system.
    The resulting distribution is virtually indistinguishable from the Poisson distribution of a perfect (e.g., lossless) photon detector, up to the threshold photon number given by the maximal number of total clicks (here, four and eight).
    To enable a fair comparison between reconstructed and ideal statistics, the mean photon number of the coherent input states was measured independently from the click-counting device.

    For determining nonclassical properties of quantum light, a thorough detector calibration and readily accessible reconstruction of quantum features, as laid out in this work, is essential.
    Thereby, experimentally accessible click-counting devices become one of the closest, readily available substitutes to true photon-number-resolving detectors.
    Furthermore, an extra application of parameters, which are commonly used to determine nonclassicality in quantum optics, was provided by exploiting them to discern fundamentally different quantum statistics in a highly sensitive manner.
    Also, the application of the binomial and Mandel parameters shows that a broad belief---a click corresponds to a photon---cannot be true because of their distinct binomial and Poissonian behavior, respectively.
    In the future, we plan to extend our framework to emerging threshold detectors, offering an intrinsic pseudo-photon-number resolution, e.g., in Refs. \cite{LMN08,SLHSSBSB23}, as well as generalization to phase-sensitive detector applications \cite{SPBTEWLNLGVASW20,PSSB24}.

\begin{acknowledgments}
	The authors acknowledge funding through the Deutsche Forschungsgemeinschaft (DFG, German Research Foundation) via the transregional collaborative research center TRR 142 (Project C10, Grant No. 231447078).
	This work received support through Ministry of Culture and Science of the State of North Rhine-Westphalia (Project PhoQC).
    This work has received funding from the German Ministry of Education and Research within the PhoQuant project (grant number 13N16103).
\end{acknowledgments}

\end{document}